\documentstyle[12pt,aasms4]{article}
\input{psfig}

\begin{document}

\title{The Cosmological Angular Momentum Problem of Low-Mass Disk Galaxies}

\author{Andreas Burkert}
\affil{Max-Planck-Institut f\"ur Astronomie, K\"onigstuhl 17,\\
       D-69117 Heidelberg, \\Germany}
\authoremail{burkert@mpia-hd.mpg.de}

\begin{abstract}
The rotational properties of the visible and dark components of low-mass
disk galaxies ($v_{rot} \leq$ 100 km/s) are investigated using the Swaters 
sample. The rotational parameter $\lambda'=\lambda (j_d/m_d)$ is determined,
where $\lambda$ is the dark halo spin parameter and $j_d/m_d$ is 
the ratio between the specific angular momentum of the disk and that
of the dark halo. The distribution of $\lambda'$ is in excellent agreement with
cosmological predictions of hierarchical clustering if $j_d/m_d \approx 1$, that is if
the protogalactic gas did not loose a significant amount of specific angular momentum due
to dynamical friction. This result is in disagreement with current
cosmological Nbody/SPH simulations where the baryonic component looses 90\% of its
specific angular momentum while settling into the equatorial plane.
The Swaters sample also shows a surprisingly strong
correlation between $\lambda'$ and the baryonic mass fraction 
$m_d$: $\lambda' = 0.4 m_d^{0.6}$. This correlation can be explained if the total amount of
gas in protogalaxies is a universal fraction of their dark matter mass, of order 10\%,
and if the variation in $m_d$ is a result of the fact that only the
inner parts of the primordial gas distribution managed to form a visible disk component.
In this case the specific angular momentum 
of the gas out of which the disk formed is a factor of 2.75 larger 
than that of the dark halo, which would require a yet unknown 
spin-up process for the visible baryonic component.

\end{abstract}


\section{Introduction}
Cosmological models of hierarchical clustering predict that galactic disks form
as a result of gas infall into cold dark matter (CDM) halos. The disk scale lengths 
and rotation curves are determined by the gravitational potential and by the 
specific angular momentum distribution which the gas acquired from tidal 
interaction with the cosmological density field (Peebles 1969) and which was modified 
during the protogalactic collapse phase.
As gas and dark matter were equally distributed initially, it is generally
assumed that both components
obtained similar amounts of specific angular momentum.
Indeed, analytical calculations (e.g. Fall \& Efstathiou 1980,
Mo, Mao \& White 1998, Mo \& Mao 2000, Buchalter et al. 2000) have shown that 
this condition can explain the observed scale lengths and various other properties 
of galactic disks, provided that the disk material retained its initial specific 
angular momentum when settling into the
galactic plane and that the dark matter halos are not too centrally condensed.

In the past couple of years detailed numerical calculations have however
uncovered serious problems with the CDM scenario.
In a seminal paper, Navarro, Frenk \& White  (1997) found that dark matter halos have
universal density profiles that can be approximated by the empirical formula
$\rho(r) = \rho_0 r_s^3/(r (r+r_s)^2)$
(hereafter the NFW profile), where $r_s = R_{200}/c$ is the scale radius, $c$ is
the concentration and $R_{200}$ is the virial radius of the dark matter halo, defined
as the radius inside which the average dark matter density is 200 times the
mean density of the universe. NFW found high concentrations of 
$c \approx 10 - 15$. Recent higher resolved simulations give 
even larger values of $c \approx 15 - 20$ 
(Moore et al. 1999, Navarro \& Steinmetz 2000).
Other simulations (e.g. by Fukushige \& Makino (1997),  Moore et al. 1998, 
Ghigna et al. 1999, Klypin et al. 2000) confirmed that DM halos are 
very centrally concentrated and found even steeper inner slopes
of $\rho \sim r^{-1.5}$. These predictions provide strong constraints for 
the slopes of galactic rotation curves and it
was noted by Flores \& Primack (1994) and Moore (1994, see also Burkert 1995;
Mc Gaugh \& de Block 1998) that NFW halos with high concentrations
disagree with the rotation curves of dark matter dominated dwarf galaxies
which rise more slowly than theoretically expected.

High-resolution cosmological  N-body/SPH simulations have also exposed the formation of galactic
disks to another problem: the scale lengths and specific angular momenta
of simulated disks are a factor of 10 smaller than observed
(Weil, Eke \& Efstathiou 1998, Navarro \& Steinmetz 1997, Sommer-Larsen, Gelato \& Vedel 1999). 
This so called {\it angular momentum
problem of galaxy formation} arises from the fact that the gas cools efficiently during the
protogalactic collapse phase, leading to dense clumps that loose a large fraction of their
angular momentum to the dark halo by dynamical friction while sinking towards the center.

Both, the halo concentration and disk angular momentum problem seem to 
provide serious challenges
for CDM and have encouraged the investigation of numerous alternative models, including 
exotic dark matter properties (e.g. Spergel \& Steinhardt 2000, Goodman 2000, Cen 2000, Peebles 2000, see however 
Dalcanton \& Hogan 2000) or the presence of a second dark baryonic spheroid 
(Burkert \& Silk 1997).  The situation has recently changed again
due to new, high-resolution observations of rotation curves.
Van den Bosch et al. (2000)  and van den Bosch \& Swaters (2000) showed
that the inner HI rotation curves of low-surface brightness (LSB) galaxies 
become significantly steeper when observed with higher 
resolution and when the effects of beam smearing are properly taken into account. 
In contrast to previous claims, the LSB rotation curves seem to be
consistent with NFW profiles, although $r^{-1.5}$-cusps might still provide
a problem.  Swaters, Madore \& Trewhella (2000) presented H$\alpha$ rotation curves
of LSB galaxies which also rise more steeply than the previously derived HI profiles and 
which are consistent with the concept of a universal rotation curve (Persic, Salucci \& Stel 1996).
In summary, at least for lower-mass galaxies the cosmologically predicted halo
profiles are not in serious
conflict with the observations anymore and the angular momentum problem has become
the main challenge for the CDM scenario.

One possibility to solve this problem is to assume energetic feedback processes which keep
the gas in a diffuse state, preventing angular momentum loss by
dynamical friction. These processes have however not proven to be very efficient.
Sommer-Larsen, Gelato \& Vedel (1999) demonstrated that the disks are still too small in simulations where
80\% or more of the final disk mass is accreted from hot, dilute gas. The situation does
only improve if the gas would be completely blown out of the galactic halos at high 
redshifts. Navarro \& Steinmetz (2000) took into account realistic feedback processes 
and found a minor impact on the angular momentum problem.

The angular momentum problem has been investigated in the past preferentially for higher
mass galaxies with rotational velocities $v_{rot} >$ 100 km/s which suffer from 
large uncertainties in the disk-halo decomposition (Syer, Mao \& Mo 1999). Recently,
van den Bosch \& Swaters (2000) published accurate disk and halo parameters for
a sample of late-type dwarf galaxies with $v_{rot} \leq$ 100 km/s.
Using their results, the rotational properties of the dark halo and the disk
in dwarf spirals are investigated and compared with theoretical predictions
in Section 2.  Section 3 shows that there exists a strong correlation between the
rotational parameter $\lambda'$ and the disk mass fraction $m_d$. A simple model of gas infall
and disk formation is presented which can explain this correlation. A discussion of
the results and conclusions follow in Section 4.

\section{The Rotational Properties of the Dark and Luminous Components in Dwarf Spiral
Galaxies}

Van den Bosch \& Swaters (2000) presented a  careful analysis of the rotation curves
of 20 late-type dwarf galaxies studied by Swaters (1999). Taking into account the
effects of beam smearing and adiabatic contraction they determined the concentration $c$
and the virial velocities $v_{200}$ of the dark
matter halos which they found to be consistent with the predictions
of $\Lambda$CDM models. In addition, the disk scale lengths $R_d$ and
the disk halo mass fractions $m_d$ were determined.
In the following we will analyse those galaxies for which
a meaningful fit could be achieved (marked by "+" in the paper by 
van den Bosch \& Swaters 2000), adopting a stellar R-band mass-to-light ratio
of $\Upsilon_R=1$.

If one assumes that the surface density distribution of the disks is exponential
and that the dark matter halos initially followed an NFW profile and responded to the growth
of the disk adiabatically, one can derive an expression for the product of the
dimensionless dark matter spin parameter $\lambda$ and the specific disk angular momentum $j_d/m_d$
(Mo, Mao \& White 1998)

\begin{equation}
\lambda' = \lambda \left(\frac{j_d}{m_d} \right) \approx \frac{\sqrt{2} R_d h}{f_R v_{200}} 
\left( \frac{2}{3} + \left( \frac{c}{21.5} \right)^{0.7} \right)^{0.5}
\end{equation}

\noindent where $h=0.75$ is the Hubble constant normalized to 100 km/s/Mpc, 
$j_d$ is the ratio of the total angular momentum of the disk
with respect to that of the dark halo and

\begin{equation}
f_R \approx ( 10 \lambda')^{-0.06+2.71m_d+0.0047/\lambda'}
\end{equation}

\noindent 
The crosses in Figure 1a show the distribution of $\lambda'$.
Unfortunately only the product of $\lambda$ and $j_d/m_d$ can be determined
directly.  It is however generally assumed that the gas initially obtained the same
specific angular momentum as the dark halo and that it conserved its
angular momentum when settling 
into the equatorial plane. In this case, $j_d/m_d = 1$ and  Figure 1a shows directly
the dark halo $\lambda$ distribution. The solid line in this figure indicates the 
theoretically predicted $\lambda$-distribution using the approximation
(Mo, Mao \& White 1998)

\begin{equation}
N(\ln(\lambda)) \sim \exp \left(- \frac{\ln^2(\lambda/\overline{\lambda})}{2 \sigma_{\lambda}^2} 
\right)
\end{equation}

\noindent with $\overline{\lambda}=0.04$ and $\sigma_{\lambda}=0.5$. The observations are in
excellent agreement with the predictions of cosmological models.
This result indicates clearly that dwarf spiral galaxies
have an angular momentum problem and that
the gas could not loose a significant amount of its angular momentum
during the protogalactic collapse phase, in contradiction to
current cosmological N-body/SPH calculations (Navarro \& Steinmetz 1997, Sommer-Larsen, Gelato \& Vedel 1999).

\section{The Correlation between $\lambda'$ and the Disk Mass Fraction $m_d$}

The data contains even more information. 
Figure 1b plots $\lambda'$ versus $m_d$ which reveals
a surprisingly strong correlation.  A least squares fit leads to

\begin{equation}
\lambda' = \lambda \left( \frac{j_d}{m_d} \right) =  0.4 m_d^{0.6}
\end{equation}

\noindent which is shown by the dashed line in Figure 1b. 

One can think of
several possibilities to explain this relationship. The sample 
of dwarf galaxies studied by van den Bosch \& Swaters (2000) could be
subject to selection effects. Figure 1a demonstrates however that
the data is not biased with respect to the dark halo spin parameter $\lambda$.
If $j_d/m_d \approx 1$ selection effects cannot affect $\lambda'$ and 
it is not clear how they could provide an explanation for equation 4.

For $(j_d/m_d) \approx 1$, equation 4 would indicate a correlation between
the halo spin parameter $\lambda$ and the disk mass fraction $m_d$.
Such a relationship is not predicted by cosmological models.

Another, maybe more promising solution is the assumption that equation 4 results from
a strong correlation between
the specific disk angular momentum $(j_d/m_d)$ and $m_d$, with the halo spin
parameter $\lambda$ being independent of $m_d$. Then, the scatter
around the dashed curve in Figure 1b reflects the cosmological scatter in 
$\lambda$ and the dashed curve represents
its mean value of $\overline{\lambda} \approx 0.04$.
With $\lambda = 0.04$, equation 4 leads to 

\begin{equation}
\frac{j_d}{m_d} = 2.5 \left( \frac{m_d}{0.1} \right)^{0.6}.
\end{equation}

There exists a simple explanation for this relationship.
Let us assume that the total gas-to-dark matter mass ratio $f_g$ in protogalaxies was universal
$f_g = M_g/M_{DM} \approx 0.1$, that however only
some fraction $m_d/f_g$ of the total gas mass, located initially
inside some radius $r_{d}$ of the protogalaxy
formed a visible disk component. If one adopts a standard top-hat model
where the distribution of gas and dark matter at the onset of the
protogalactic collapse phase can be approximated by
a homogeneous, rigidly rotating sphere with total radius R, the specific angular momentum
of the gas inside $r_d \leq$ R is $(J/M)_g=0.4 \omega_g r_d^2$ where $\omega_g$ is the
angular velocity of the gas sphere.  The specific angular momentum of the dark component
is given by $(J/M)_{DM}=0.4 \omega_{DM} R^2$ with $\omega_{DM}$ the specific angular
velocity of the dark matter. As the mass fraction of gas inside $r_d$ is
$m_d/f_g=(r_d/R)^3$ we find 

\begin{equation}
\left( \frac{j_d}{m_d} \right) = \frac{(J/M)_g}{(J/M)_{DM}} = 
 \frac{\omega_g}{\omega_{DM}} 
\left( \frac{m_d}{f_g} \right)^{2/3}
\end{equation}

\noindent which has the same power-law dependence as equation 5.

\section{Discussion of the Results and Conclusions}

The previous analysis leads to some interesting conclusions which might be useful for further
studies of the cosmological angular momentum problem.

The observed disk mass ratios are
$m_d \leq 0.1$ which implies that this value is a lower 
limit for the universal primordial gas fraction
in protogalaxies, i.e. $f_g \geq 0.1$. In many dwarf spiral galaxies $m_d$  
is much smaller than 0.1.
Most of their primordial gas might therefore not have
contributed to the visible disk component.
This gas could still be around in a diffuse state or
it could have been blown out by galactic winds.

Adopting $\lambda = 0.04$,
a good fit to the observed correlation (Figure 1b) can be achieved if

\begin{equation}
\frac{\lambda'}{0.04} = j_d/m_d = 2.75 (m_d/0.1)^{2/3}.
\end{equation}

\noindent This relationship is shown by the dotted line in Fig. 1b.
Comparing equation 7 with equation 6 leads to

\begin{equation}
\omega_g/\omega_{DM} \approx 2.75 (f_g/0.1)^{2/3}.
\end{equation}

\noindent As $f_g \geq 0.1$  
the gas out of which the galactic disks formed either was 
rotating faster than the dark halos ($\omega_g \geq 2.75 \omega_{DM}$) initially
or it gained angular momentum lateron.
Note that all galaxies in the present sample must have experienced the same spin-up
event in order to explain the strong correlation observed.

Up to now, it is not clear which processes
could  have increased the specific angular momentum of the gaseous components
beyond that of the dark halos.
One possibility might be selective mass loss
in galactic winds which might have ejected preferentially the low angular momentum gas,
leaving behind  disks with high specific angular momentum (van den Bosch, private
communication). Although this scenario is interesting it is not clear
how the angular momentum - mass relationship (equation 4) could be explained
within the framework of this model. In addition, galaxies with large values of
$m_d=0.1 \approx f_g$ probably did not loose a significant amount of gas.
Yet they have the highest values of $\lambda' = 0.1$ and therefore the highest
specific angular momenta of all dwarf spirals.

\acknowledgments

I thank the staff of Kyoto University for their hospitality
during my visit on which this work was started, Frank van den Bosch for a careful reading
of the manuscript and important suggestions and Houjun Mo for interesting discussions.

\newpage

\begin{figure}
\plotone{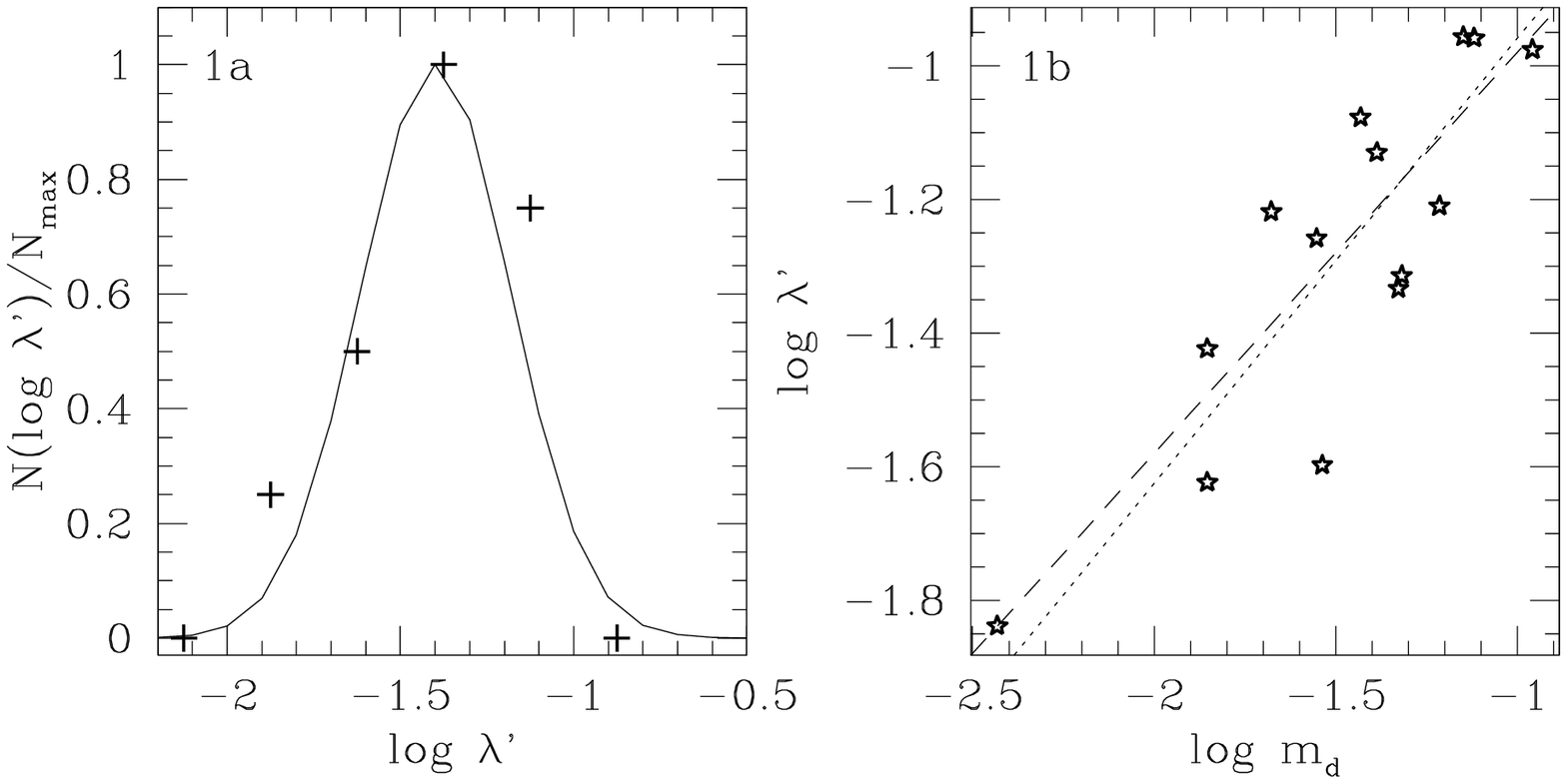}
\figcaption{
Figure 1a compares the distribution of $\lambda'=\lambda(j_d/m_d)$ of the Swaters sample (crosses)
with the predictions of cosmological models adopting $(j_d/m_d)=1$ (solid line). Figure 1b shows
as open starred symbols the correlation between
$\lambda'$ and the disk mass fraction $m_d$ for the Swaters sample. The dashed curve represents
a linear least squares fit through the data. The dotted line shows the predicted relationship
for the model discussed in the text.}
\end{figure}


\begin{references}
\reference{bu00} Buchalter, A., Jimenez, R. \& Kamionkowski, M. 2000, \mnras, submitted, astro-ph/0006032
\reference{bu95} Burkert, A. 1995, \apj, 447, L25
\reference{bu97} Burkert, A. \& Silk, J. 1997, \apj, 488, L55
\reference{ce00} Cen, R. 2000, \apj, submitted, astro-ph/0005206
\reference{da00} Dalcanton, J.J. \& Hogan, C.J. 2000, \apj, submitted, astro-ph/0004115
\reference{fa80} Fall, S.M. \& Efstathiou, G. 1980, \apj, 193, 189
\reference{fl94} Flores, R.A. \& Primack, J.R. 1994, \apj, 427, L1
\reference{fa80} Fukushige, T. \& Makino, J. 1997, \apj, 477, L9
\reference{gh99} Ghigna, S., Moore, B., Governato, F., Lake, G., Quinn, T. \& Stadel, J. 1999,
\apj, submitted, astro-ph/9910166
\reference{go00} Goodman, J. 2000, \apj, submitted, astro-ph/0003018
\reference{kl00} Klypin, A., Kravtsov, A.V., Bullock, J.S. \& Primack, J.R. 2000, \apj, submitted,
astro-ph/0006343
\reference{mc98} McGaugh, S.S. \& de Blok, W.J.G. 1998, \apj, 499, 41
\reference{mo98} Mo, H.J., Mao, S. \& White, S.D.M. 1998, \mnras, 295, 319
\reference{mo00} Mo, H.J. \& Mao, S. 2000, \mnras, in press, astro-ph/0002451
\reference{moo94} Moore, B. 1994, Nature, 370, 629
\reference{moo98} Moore, B., Governato, F., Quinn, T., Stadel, J.
\& Lake, G. 1998, \apj, 499, L5
\reference{mo99} Moore, B., Ghigna, S., Governato, F., Lake, G., Quinn, T., Stadel, J.
\& Tozzi, P. 1999, \apj, 524, L19
\reference{na97} Navarro, J., Frenk, C.S. \& White, S.D.M. 1997, \apj, 490, 493
\reference{nav97} Navarro, J. \& Steinmetz, M. 1997, \apj, 478, 13
\reference{na00} Navarro, J. \& Steinmetz, M. 2000, \apj, 528, 607
\reference{pe69} Peebles, P.J.E. 1969, \apj, 155, 393
\reference{pe00} Peebles, P.J.E. 2000, \apj, 528, L61
\reference{pe96} Persic, M., Salucci, P. \& Stel, F. 1996, \mnras, 281, 27
\reference{so99} Sommer-Larsen, J., Gelato, S. \& Vedel, H. 1999, \apj, 519, 501
\reference{sp00} Spergel, D.N. \& Steinhardt, P.J. 2000, Phys. Rev. Lett., 84, 3760
\reference{sw00} Swaters, R.A., Madore, B.F. \& Trewhella, M. 2000, \apj, 531, L107
\reference{sw99} Swaters, R.A. 1999, PhD Thesis, University of Groningen.
\reference{sy00} Syer, D., Mao, S. \& Mo, H.J. 1999, \mnras, 305, 357
\reference{va00} van den Bosch, F.C., Robertson, B.E., Dalcanton, J.J. \& de Blok, W.J.G. 2000,
\aj, 119, 1579
\reference{van00} van den Bosch, F.C. \& Swaters, R.A. 2000, \aj, submitted, astro-ph/0006048
\reference{we98} Weil, M.L., Eke, V.R. \& Efstathiou, G. 1998, \mnras, 300, 773
\end{references}
\end{document}